# L-band radiometric behaviour of pine forests for a variety of surface moisture conditions

J.P. Grant (1,2), J.-P. Wigneron (2), A.A. Van de Griend (1), F. Demontoux (3), G. Ruffié (3), A. Della Vecchia (4), N. Skou (5), B. Le Crom (3)
*(1) Vrije Universiteit Amsterdam, Dept. Hydrology & Geo-environmental Sciences, De Boelelaan 1085, 1081 HV, Amsterdam, The Netherlands, (2) INRA, EPHYSE, Bordeaux, France, (3) IMS (PIOM), Université de Bordeaux, France, (4) Universita di Roma Tor Vergata, Italy, (5) Ørsted-DTU, Technical University of Denmark*
jennifer.grant@falw.vu.nl

ABSTRACT - *From July-December 2004 the experimental campaign 'Bray 2004' was conducted in the coniferous forest of Les Landes near Bordeaux, France, using a multi-angle L-band (1.4 GHz) radiometer to measure upwelling radiation above the forest. At the same time, ground measurements were taken of soil and litter moisture content. This experiment was done in the context of the upcoming SMOS mission in order to improve our understanding of the behaviour of the L-band signal above forested areas. Very little information exists on this subject at the moment, especially for varying hydrological conditions. Furthermore, additional measurements were done at the University of Bordeaux (IMS laboratory) to determine the dielectric behaviour of a litter layer such as that found at the Bray site. There is some evidence that this layer may have a different influence on the L-band signal than either the soil or the vegetation, however the exact behaviour of the litter layer and the extent of its influence on the L-band signal are as yet unknown. This paper presents 1) results of the Bray experiment describing the behaviour of the above-canopy L-band emissivity for different conditions of ground moisture and 2) the relationship between soil and litter moisture content and results of the laboratory experiments on litter dielectric properties. Together this will give a first insight into the L-band radiometric properties of the different forest layers for varying hydrological conditions.*

## 1 INTRODUCTION

Soil moisture is a key variable controlling the exchange of heat and moisture between the land and the atmosphere through evaporation. There is currently a lack of global soil moisture observations, which are necessary to improve our knowledge of the water cycle and to contribute to better weather and climate forecasting. For this reason the European Space Agency (ESA) has developed the Soil Moisture and Ocean Salinity (SMOS) mission, to be launched in 2008, as part of its Living Planet Programme (e.g. Kerr *et al.*, 2001).

SMOS will carry a dual-polarization, multi-angle (0º-55º) L-band radiometer and provide maps of surface soil moisture over land surfaces and salinity over the oceans. Temporal resolution will be 2-3 days, and spatial resolution will be around 40 km at nadir. At this spatial resolution, many land surface pixels will be inhomogeneous. Few pixels will contain 100% bare soil, therefore the influence of vegetation on the signal should be accounted for. A vegetation layer will attenuate the soil emission and add its own emission to the signal, an effect which increases under wet conditions. Most studies on this subject have focussed on crops. However, a large amount of SMOS pixels will also contain partial forest cover and at present there is little existing knowledge of the influence of this vegetation type on the L-band signal. Most studies on L-band forest radiometry are based on modelling or very short-term field observations (e.g. Lang *et al.* (2001); Ferrazzoli *et al.* (2002); Saleh *et al.* (2004); Della Vecchia *et al.* (2006)). This was the reason to conduct the long-term field experiment 'Bray 2004' over a pine forest and study the effect of the forest on the L-band signal under varying surface moisture conditions (Grant *et al.*, 2006).

From the literature comes increasing evidence that a litter layer will also contribute substantially to the above-canopy emission (Schmugge *et al.*, 1988; Jackson & Schmugge, 1991; Saleh *et al.*, 2006). Therefore, Bray soil and litter dielectric properties were measured at the IMS/PIOM Laboratory, in order to model the resulting emissivity of a soil-litter system (Le Crom *et al.*, 2006).

The objective of this study is *first*, to describe the L-band signal above forests for varying surface moisture conditions and *second*, to present the results of litter permittivity measurements. This will give a first insight into the properties of the different forest layers at L-band for varying hydrological conditions. The resulting information will eventually be incorporated into the forward model of the SMOS Level 2 algorithm: L-band Microwave Emission of the Biosphere (L-MEB) (Wigneron *et al*., 2006).

## 2 METHODS AND MATERIALS

### 2.1 Site description

The Bray site lies within the forest of Les Landes, southwest of Bordeaux, France (latitude 44°42′ N, longitude 0°46′ W, altitude 61 m). Les Landes forest is a production forest consisting mainly of Maritime Pines (*Pinus pinaster* Ait). The trees at the Bray site were 34 years in age at the time of measurement, giving the stand an approximate height of 22 m. The trees are distributed in parallel rows along a northeast-southwest axis with an inter-row spacing of 4 m. Maximum (summer) values for canopy LAI and cover fraction were around 2.15 and 0.35 respectively (from measurements by INRA-Bordeaux). The understory consists mostly of grass (mainly *Molinia caerulea* L. Moench) and had maximum LAI and cover fraction values of around 2.48 and 0.65 respectively (from measurements by INRA-Bordeaux).

The soils are sandy and hydromorphic podzols, with dark organic matter in the first 60 cm. The percentage of sand in the soil surface layer generally exceeds 80%. On top of the soil lies a distinct litter layer, the upper part of which consists mainly of dead grass and the lower part of grass roots, pine needles and other organic matter. In places the layer thickness exceeded 10 cm, and the large biomass was also indicated by measurements of water content resulting in values of over 10 kgm$^{-2}$.

### 2.2 Measurements

#### 2.2.1 Remote Sensing

Microwave measurements were done with the dual-polarization L-band (1.41 GHz) radiometer EMIRAD of the Technical University of Denmark, of which technical details can be found in (Søbjærg, 2002).

At the Bray site, the radiometer was mounted on a 40 m mast over the forest, giving it a footprint of approximately 600 m$^2$ at an incidence angle of 45°. Measurements were done automatically at incidence angles of 25°, 30°, 35°, 40°, 45°, 50°, 55° and 60° from nadir and averaged to half-hourly values for the final data analysis. A sky calibration was done at intervals throughout the six-month period. Only horizontally polarized measurements were available for this experiment. Full experimental details can be found in (Grant *et al*., 2006).

A thermal infrared (IR) radiometer (Heitronics KT 15.85D; 9.6 – 11.5 μm) was fixed next to the microwave instrument to give measurements of surface temperature over approximately the same footprint.

#### 2.2.2 Field measurements

Soil temperature was measured at four different locations at depths of 1, 2, 4, 8, 16, 32, 64 and 100 cm below the soil surface, using thermocouples made by INRA and a CR21X Campbell Scientific data logger. Temperature measurements were taken every 10 s and averaged to half-hourly values. The same method was used to record litter temperatures at 1, 3 and 5 cm above the mineral soil surface.

For the measurements of soil and litter moisture content, 3 ThetaProbes (Delta-T Devices Ltd., type ML2x) were placed in the soil layer and 3 in the litter layer. The ThetaProbes each consisted of 4 rods of 60 mm length, which were placed in the respective layer at an angle of approximately 20°. The probes were connected to a CS21X Campbell Scientific data logger, which averaged the measurements taken every 10 seconds to give half hourly values. Periodic soil and litter samples were taken at random locations at the site for calibration purposes. Dry bulk density of the Bray soil was 1.25 gcm$^{-3}$ from previous experiments. Again, full experimental details can be found in (Grant *et al*., 2006).

N.B. Soil and litter moisture contents are given in volumetric and gravimetric percentages respectively. Unit conversion is as follows: 1 % = 1 m$^3$m$^{-3}$.

#### 2.2.3 Laboratory measurements (litter)

After taking soil and litter samples at the Bray site, laboratory measurements of the dielectric properties of the soil and litter were done using a wave guide technique. This method enabled the use of samples wide enough to account for the layer heterogeneity. Sample thickness was 1 cm for soil and 2 cm for litter. Waveguide dimensions were 129.27 x 54.77 mm. The samples were held inside the guide using a support with as a base a 100 μm thick Mylar sheet, considered to be quasi-transparent for the electromagnetic waves. The electromagnetic parameters of the samples were determined using the Nicolson Ross Weir method (NRW) for rectangular waveguides. The principle of the calculation is based on the fact that introduction of the sample into the guide produces a change of characteristic impedance. Full experimental & modeling details can be found in (Le Crom *et al*., 2006).

## 2.3 Calculations

Emissivity calculations were based on the Rayleigh-Jeans approximation for the microwave domain:

$$e_{\text{surf}}(\theta, P) = T_B(\theta, P) / T_{\text{surf}} \quad (1)$$

where $T_B$ is the observed brightness temperature, $\theta$ and $P$ are the incidence angle and polarization of the measurement, respectively, and $T_{\text{surf}}$ and $e_{\text{surf}}$ are the temperature and emissivity, respectively, of the emitting surface. Emissivity is a function of soil moisture, which relationship can be described with a dielectric mixing model (the method of Dobson *et al.* (1985) is used here and in L-MEB) and the Fresnel equations.

L-MEB includes a calculation for an effective ground-canopy temperature (Wigneron *et al.*, 2006), which was used here to find $T_{\text{surf}}$:

$$T_{\text{surf}} = A_t \cdot T_{\text{canopy}} + (1 - A_t) \cdot T_{\text{soil}} \quad (2)$$

with ($0 \leq A_t \leq 1$):

$$A_t = B_t \cdot (1 - \Gamma(\theta, P)) \quad (3)$$

where $B_t$ is a canopy type-dependent parameter, and:

$$\Gamma(\theta, P) = e^{-\tau_0 / \cos\theta} \quad (4)$$

where $\Gamma(\theta, P)$ is the canopy transmissivity and $\tau_0$ is the vegetation optical depth at nadir.

In this study, the effective soil temperature $T_{\text{soil}}$ was calculated according to the method described in (Ulaby *et al.*, 1986), using soil temperature and dielectric (i.e. moisture) profiles. Canopy temperature $T_{\text{canopy}}$ was taken from the IR measurements, which were found to show 96% correlation with branch temperatures measured at Bray. $A_t$ was found by optimizing for $B_t$ and $\tau_0$, which resulted in $B_t = 0.49 \pm 0.13$ and $\tau_0 = 0.62 \pm 0.24$.

## 2.4 Modelling

Modelling was done using the L-MEB model (Wigneron *et al.*, 2006), which for a vegetation-covered surface is based on a simplified radiative transfer model, also known as the τ-ω model:

$$T_B = T_g e_g \Gamma + T_v (1 - \omega)(1 - \Gamma)(1 + \Gamma \cdot (1 - e_g)) \quad (5)$$

where the subscripts 'g' and 'v' denote ground and vegetation, respectively, and $\omega$ is the single scattering albedo of the vegetation canopy.

This model accounts for *1)* direct vegetation emission, *2)* soil emission attenuated by the canopy and *3)* vegetation emission reflected by the soil and attenuated by the canopy.

## 3 RESULTS AND DISCUSSION

### 3.1 Temperature and moisture

Correlation between measured brightness temperatures and the effective ground-canopy temperature was found to be 85%. Figure 1 gives a visual example of this relationship for the period Julian Day (JD) 271-276. The high correlation indicates that in this case much of the horizontally polarized L-band signal is dominated by temperature influences.

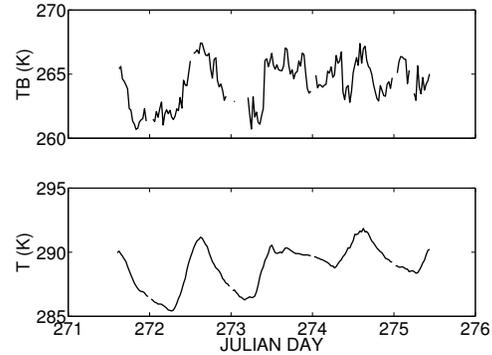

**Figure 1**: Measured brightness temperatures $T_B$ (top) and calculated surface temperature $T_{\text{surf}}$ (bottom) for the period JD 271-276; $R^2 = 0.85$.

Figure 2 shows brightness temperatures plotted against incidence angle for 'wet' (soil moisture SM > 25%) and 'dry' (SM < 15% and no precipitation). There is a clear difference in 'angular signal' for both conditions, showing that multi-angular measurements, such as those of SMOS, contain information on surface moisture conditions.

The pattern of a decreasing emissivity with increasing viewing angle is a typical soil pattern, whereas a typical canopy pattern shows less angular influence. Soil emission decreases under wet conditions, whereas canopy emission increases. A wet bare soil shows a higher range and lower values of emissivity compared to a dry bare soil, and the patterns shown in figure 2 are therefore not unexpected. However, the possible influence of a litter layer on the above-canopy signal should still be considered and investigated.

A strong relation between soil and litter moisture was found at this site (figure 3), thus making it difficult to decouple the effects of soil and litter moisture on the above-canopy signal.

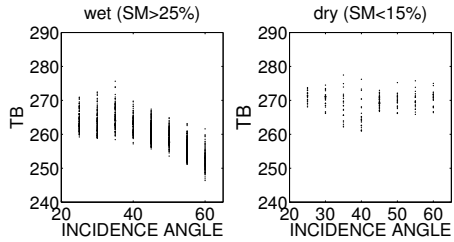

**Figure 2**: Measured brightness temperatures $T_B$ for incidence angles 25°-60°. Left: 'wet' measurements at times when SM > 25%; right: 'dry' measurements at times when SM < 15% and no precipitation.

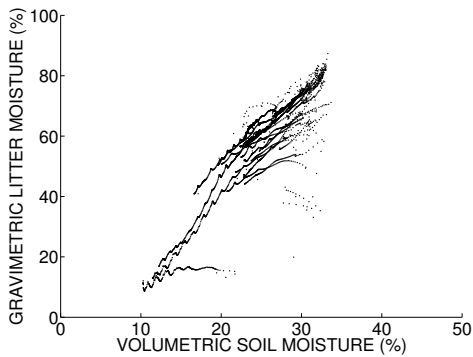

**Figure 3**: Soil moisture (vol.%) *vs.* litter moisture (grav.%) for the Bray site; $R^2 = 0.84$.

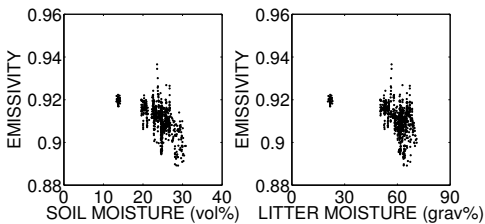

**Figure 4**: Emissivity calculated from brightness temperature measurements *vs.* soil (left) and litter (right) moisture content.

When above-canopy emissivity is plotted against either soil or litter moisture, as in figure 4 (45° measurements only), a very small dynamic range is found: ~ 0.04 change in emissivity for a ~ 20% range in soil moisture or a ~ 60% range in litter moisture (similar ranges; see fig.3). Average emissivity values are high. This shows that a forest system such as that found at Bray has a very low sensitivity to variations in soil moisture and it is therefore doubtful whether soil moisture content can be retrieved with a meaningful precision in this kind of environment.

3.2 Litter

Figures 5 and 6 show the results of permittivity measurements of soil and litter, respectively. Only the real part ($\varepsilon'$) of the dielectric constant is shown here, as at 1.4 GHz this is the main part of the dielectric constant affecting emissivity. The figures give a band of permittivity values, in order to include the effects of medium heterogeneity and errors of measurement. A detailed explanation and further measurements can be found in (Le Crom *et al.*, 2006).

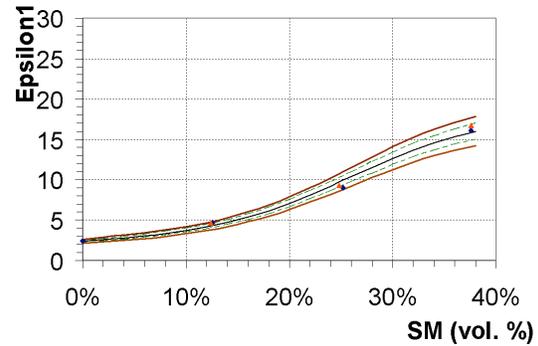

**Figure 5**: Field of soil permittivity ($\varepsilon'$) *vs.* moisture content.

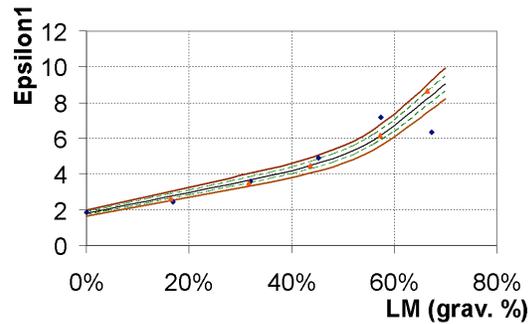

**Figure 6**: Field of litter permittivity ($\varepsilon'$) *vs.* moisture content.

The figures show that for similar moisture ranges (from fig. 3), the range in soil permittivity is twice that of litter. Therefore, if a substantial litter layer is present at a given site, but ignored, this could result in a severe underestimation of soil moisture content at higher wetness conditions (from a dielectric mixing model (e.g. Dobson *et al*., 1985) and the Fresnel equations).

From figures 5 and 6, and the relationship given in fig. 3, it was possible to model the emissivity of a soil overlain by a 3 cm litter layer, as a function of soil moisture content (Le Crom *et al*., 2006). The result is shown in figure 7. This information will be used for future evaluation and/or adaptation of the L-MEB model to account for the effect of a litter layer on the L-band signal.

Using L-MEB (eq. 5) and assumed values of $\Gamma = 0.39$ (at a 45º angle) and $\omega = 0.08$ for the Bray canopy (from Della Vecchia and Ferrazzoli, 2006), we calculated the ground emissivity, which in this case is the soil+litter emissivity.

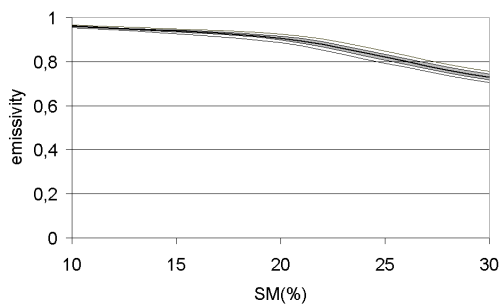

**Figure 7**: Field of emissivity as a function of moisture content for a soil overlain by a 3 cm litter layer.

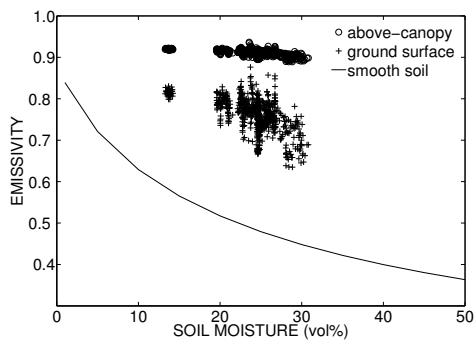

**Figure 8**: Comparison of above-canopy, ground surface and smooth soil emissivities as a function of soil moisture content.

Results are shown in figure 8, where above-canopy (fig. 4), ground surface (soil+litter system) and smooth soil (from Dobson *et al*. (1985)) emissivities are compared as a function of soil moisture content. Theoretically, the ground surface emissivity in this figure should be the same as the result given in figure 7. However, probably due to the use of different models and differences in layer thickness, the results are not exactly the same, although the magnitude of the range is similar. The figures should therefore be taken as a first indication of the emissivity ranges involved.

4 SUMMARY AND CONCLUSION

The greater part of the horizontally polarized L-band signal is dominated by temperature influences. Variations in soil and/or litter moisture are visible in the angular signal and in the above-canopy microwave emission, although the dynamic range of this last effect is very small. This, together with the fact that emissivity values are very high, is possibly due to the presence of a substantial litter layer.

Decoupling of soil and litter effects is difficult because of a strong correlation between soil and litter moisture. Therefore, laboratory measurements and modelling were done at IMS laboratory to improve our understanding of this issue. For similar moisture ranges, the range in soil permittivity was found to be twice that of litter. Ignoring the presence of a litter layer could therefore result in a severe underestimation of soil moisture content at higher wetness conditions.

Results of these studies will be used for future evaluation/adaptation of the L-MEB model used for SMOS.

5 REFERENCES

Dobson, M.C., Ulaby, F.T., Hallikainen, M.T. and El-Rayes, M.A., 1985, Microwave Dielectric Behaviour of Wet Soil – Part II: Dielectric Mixing Models. *IEEE Transactions on Geoscience and Remote Sensing*, 23, 35-46.

Della Vecchia, A. and Ferrazzoli, P., 2006, A Large-Scale Approach to Estimate L-band Emission from Forest Covered Surfaces. ESA Technical Note SO-TN-TV-GS-0001-01.a, SMPPD Study, February 2006.

Della Vecchia, A., Saleh, K., Ferrazzoli, P., Gurriero, L and Wigneron, J.-P., 2006, Simulating L-band Emission of Coniferous Forests Using a Discrete Model and a Detailed Geometrical Representation. *IEEE Transactions on Geoscience and Remote Sensing Letters*, 3(3), 364-368.

Ferrazzoli, P., Gurriero, L. and Wigneron, J.-P., 2002, Simulating L-band Emission of Forests in View of Future Satellite Applications. *IEEE Transactions on Geoscience and Remote Sensing*, 40(12), 2700-2708.

Grant, J.P., Wigneron, J.-P., Van de Griend, A.A., Schmidl Søbjærg, S. and Skou, N., 2006, First results of the 'Bray 2004' field experiment on L-band forest radiometry – microwave signal behaviour for varying conditions of surface moisture. *Remote Sensing of Environment*, submitted.

Jackson, T.J. and Schmugge, T.J., 1991, Vegetation Effects on the Microwave Emission of Soils. *Remote Sensing of Environment*, 36, 203-212.

Kerr, Y.H., Waldteufel, P., Wigneron, J.-P., Font, J. and Berger, M., 2001, Soil Moisture Retrieval from Space: The Soil Moisture and Ocean Salinity (SMOS) Mission. *IEEE Transactions on Geoscience and Remote Sensing.*, 39(8), 1729-1735.

Lang, R.H., Utku, C., De Matthaeis, P., Chauban, N. and Le Vine, D.M., 2001, ESTAR and Model Brightness Temperatures over Forests: Effects of soil moisture. Proceedings of IGARSS-01, Sydney.

Le Crom, B., Demontoux, F., Ruffié, G., Wigneron, J.-P. and Grant, J.P., 2006, Electromagnetic Characterization of Soil-Litter Media, Application to the simulation of the microwave emissivity of the ground surface in forests. *IEEE Transactions on Geoscience and Remote Sensing*, in preparation.

Saleh, K., Wigneron, J.-P., Calvet, J.-C., Lopez-Baeza, E., Ferrazzoli, P., Berger, M., Wursteisen, P., SImmonds, L. and Miller, J., 2004, The EuroSTARRS Airborne Campaign in Support of the SMOS Mission: First results over land surfaces. *International Journal of Remote Sensing*, 25(1), 177-194.

Saleh, K., Wigneron, J.-P., De Rosnay, P., Calvet, J.-C., Escorihuela, M.J., Kerr, Y. and Waldteufel, P., 2006, Impact of Rain Interception by Vegetation and Mulch on the L-band Emission of Natural Grass (SMOSREX Experiment). *Remote Sensing of Environment*, 101(1), 127-139.

Schmugge, T.J., Wang, J. R. and Asrar, G., 1988, Results from the Push Broom Microwave Radiometer Flights over the Konza Prairie in 1985. *IEEE Transactions on Geoscience and Remote Sensing*, 26(5), 590-596.

Søbjærg Schmidl, S., 2002, Polarimetric Radiometers and their Applications. PhD Thesis, Technical University of Denmark, 144 pages.

Ulaby, F., Moore, and Fung, A., 1986, Microwave Remote Sensing: Active and Passive, Vol. III: From theory to applications, Artech House, Dedham, MA. 1537-1539.

Wigneron, J.-P., Kerr, Y., Waldteufel, P., Saleh, K., Richaume, P., Ferrazzoli, P., Escorihuela, M.-J., Grant, J.P., Hornbuckle, B., De Rosnay, P., Calvet, J.-C., Pellarin, T., Gurney, R. and Mätzler, C., 2006, L-band Microwave Emission of the Biosphere (L-MEB) Model, Results from calibration against experimental data sets over crop fields. *Remote Sensing of Environment*, in press.